%% file: conference.tex
\documentclass[sigconf]{acmart}
\usepackage{hyperref,amsmath,amssymb,amsfonts}
\usepackage{algorithmic}
\usepackage{graphicx}
\usepackage{textcomp}
\usepackage{balance}
\usepackage{xcolor}
\usepackage{acronym}
\usepackage{booktabs,tabularx}
\usepackage{subfig}
\usepackage{commath}
\input{./00_acronyms.tex} 		

\begin{document}
\acmConference[WI’19]{The 18th Web Intelligence conference}
\acmBooktitle{The 18th Web Intelligence conference.}
\fancyhead{}

\title[\acs{P2PL} recommendation engine]{Recommendation Engine for Lower
  Interest Borrowing on Peer to Peer Lending (P2PL) Platform}

\author{Ke Ren}
\affiliation{%
	\institution{Department of Electrical and Computer Engineering, The University of Auckland, NZ}
}
\email{keith\_kt@foxmail.com}
\author{Avinash Malik}
\affiliation{%
	\institution{Department of Electrical and Computer Engineering, The University of Auckland, NZ}
}
\email{avinash.malik@auckland.ac.nz}
 
\begin{abstract}
  Online \ac{P2PL} systems connect lenders and borrowers directly,
  thereby making it convenient to borrow and lend money without
  intermediaries such as banks. Many recommendation systems have been
  developed for lenders to achieve higher interest rates and avoid
  defaulting loans. However, there has not been much research in
  developing recommendation systems to help borrowers make wise
  decisions.  On \ac{P2PL} platforms, borrowers can either apply for
  bidding loans, where the interest rate is determined by lenders
  bidding on a loan or traditional loans where the \ac{P2PL} platform
  determines the interest rate. Different borrower \textit{grades} ---
  determining the credit worthiness of borrowers get different interest
  rates via these two mechanisms.  Hence, it is essential to determine
  which type of loans borrowers should apply for. In this paper, we
  build a recommendation system that recommends to any new borrower the
  type of loan they should apply for.  Using our recommendation system,
  any borrower can achieve lowered interest rates with a higher
  likelihood of getting funded.
\end{abstract}
\maketitle

\section{Introduction}
\label{sec:introduction}

The development of electronic commerce has lead to a burgeoning growth
in online \acf{P2PL} system. \ac{P2PL} system is a micro financing
platform, which is rising as an alternative to traditional financial
lenders such as banks. There are two main participants in \ac{P2PL}
systems: borrowers and lenders. On the one side, borrowers apply for
loans. On the other side, lenders can view the characteristics of the
borrowers/loans and decide, which loans they should invest in. In recent
years, a great deal of research has gone into developing recommendation
systems to help lenders~\cite{Guo2016417, ren2019investment} achieve
high returns with low risk of defaults. However, there has not been much
research into developing recommendation systems to advice borrowers. In
particular, the main objective from borrower's perspective is getting
funded with the lowest interest rate payable. We build a recommendation
framework for borrowers to help them borrow with lower interest rates
and increased likelihood of getting funded on \ac{P2PL} platforms in
this paper.

\begin{table}[t!]
  \caption{Average interest rates of traditional loans and bidding loans
    for borrowers with the same characteristics. }
  \begin{center}
    \resizebox{0.8\columnwidth}{!}{%
      \begin{tabular}{cccccccc}\toprule
        \hline
        Grade            & AA   & A    & B    & C    & D    & E    & HR \\ \midrule
        Average traditional interest & 0.112 & 0.082 & 0.158 & 0.197 & 0.247 & 0.295 & 0.318 \\ 
        Average bidding interest     & 0.113 & 0.102 & 0.151 & 0.182 & 0.208 & 0.247 & 0.235 \\ 
        Traditional $-$ bidding     & -0.001 & -0.02 & 0.007 & 0.015 & 0.039 & 0.048 & 0.083 \\ \bottomrule
      \end{tabular}%
    }
    \label{tab:1}
  \end{center}
\end{table}

\begin{table}[t!]
  \caption{T-test between the interest rates of traditional loans and
    bidding loans for each grade, with null hypothesis that they have
    the same mean value. }
  \begin{center}
    \resizebox{0.8\columnwidth}{!}{%
      \begin{tabular}{cccccccc}\toprule
        \hline
        Grade            & AA   & A    & B    & C    & D    & E    & HR \\ \midrule
        P-value  & 0.74 & 3.6e-09 & 0.061 & 1.41e-05 & 1.15e-22 & 1.40e-27 & 1.77e-29 \\ 
        Decision & Not reject & Reject & Not reject & Reject & Reject & Reject & Reject \\ \bottomrule
      \end{tabular}%
    }
    \label{tab:t-test}
  \end{center}
\end{table}

From the borrower's perspective, there are two essential questions that
need to be considered when applying for loans: \textcircled{1} will the
loan be funded successfully? \textcircled{2} What is the lowest
obtainable interest rate? Online \ac{P2PL} platforms do not help
borrowers with these two questions, but rather give the borrowers a
choice to select from different types of loans that they can apply
for. On online \ac{P2PL} platforms\footnote{prosper.com}, the two main
types of loans are:

\begin{itemize}
\item \textbf{Traditional loan:} based on the borrower's personal
  information, \ac{P2PL} platforms decide the interest rate for each
  borrower's loan. Next, the \ac{P2PL} platforms put the loan online for
  a certain period for lenders to fund the loan.
\item \textbf{Bidding loan:} first and foremost, borrowers themselves
  decide the maximum interest rate they are willing to pay. Then
  \ac{P2PL} platforms put the loan online and wait for lenders to bid on
  the loan, with the interest rate that they want.  At the end of the
  bidding period, if the loan receives sufficient funding, \ac{P2PL}
  platforms will select lenders with the lowest interest rate. However,
  if the final total interest rate is higher than the borrower's maximum
  selected interest rate, then this loan is \textit{not} funded.
\end{itemize}

\begin{figure}[!t]
  \centering
  \includegraphics[width=2in]{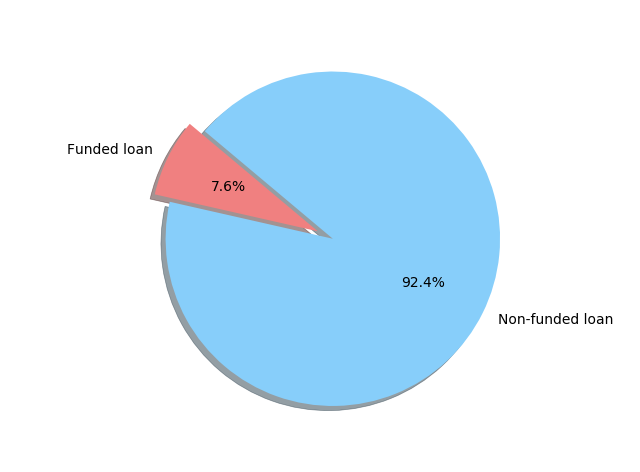}
  \caption{Pie chart of distribution on 12006 bidding loans}
  \label{fig:pie}
\end{figure}

To show that it is necessary to select the right type of loan when
applying on \ac{P2PL} platforms, consider the average historical
interest rate for traditional and bidding loans from Prosper (one of the
largest \ac{P2PL} platform in the world) in Table~\ref{tab:1}.
Table~\ref{tab:1} shows the interest rates of traditional and bidding
loans for each borrower grade along with their differences. A higher
grade (e.g., AA) indicates lower likelihood of the borrower defaulting
and a lower grade (e.g., HR) indicates higher likelihood of the borrower
defaulting on their loan obligations.  The T-test with the
null-hypothesis that the traditional and bidding loans have the same
mean value is shown in Table~\ref{tab:t-test}.  We can observe from
Tables~\ref{tab:1} and~\ref{tab:t-test} that borrowers with credit grade
A should apply for a traditional loan, while borrowers with lower credit
grades C, D, E, and HR would achieve a lower interest rate payable when
applying for bidding loans. Finally, borrowers with credit grade AA and
B can either apply for bidding or traditional loans, since there is no
significant statistical difference between the interest rates of bidding
and traditional loans, for these grades.  Especially for borrowers with
HR grade, the interest rate payable, when applying for a bidding loan,
is decreased by 8.3\%. Hence, it is necessary for borrowers to decide,
which types of loan should they apply for.

Getting a lower interest rate is one borrower objective, the other
objective is to actually get funded. Figure~\ref{fig:pie} illustrates
the distribution of funded and non-funded bidding loans from a total of
12006 loans from the Prosper historical dataset. It can be seen that on
\textit{average} only 7.6\% of all bidding loans get funded. The success
of getting funded for different grades of bidding loans is shown in
Table~\ref{tab:successrategradesbidding}. Thereby making it important
for borrowers to make a wise choice when applying for a loan. 

\begin{table}[tb]
  \centering
  \begin{tabular}{|c|c|c|c|c|c|c|}
    \hline
    AA&A&B&C&D&E&HR \\
    \hline
    34.2\% & 33.1\% & 27.3\% & 16.1\% & 10.4\% & 4.7\% & 1.6\%\\
    \hline
  \end{tabular}
  \caption{Average success rate of funding bidding loans of different
    grades.}
  \label{tab:successrategradesbidding}
\end{table}

\begin{figure}
  \centering
  \includegraphics[width=3in]{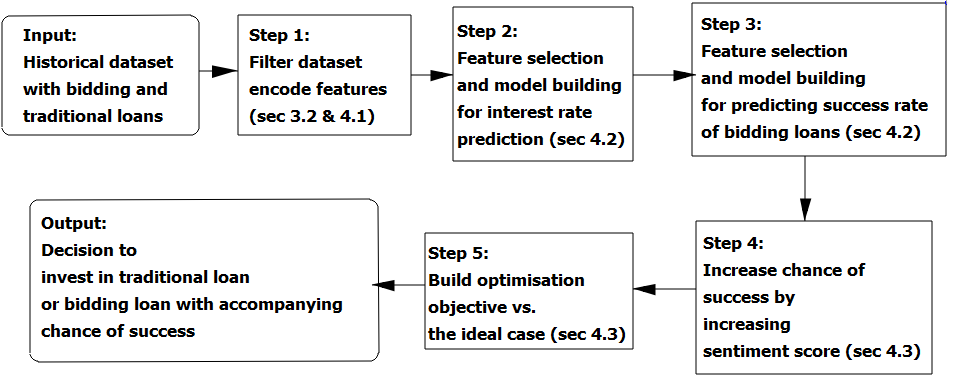}
  \caption{Overall proposed methodology}
  \label{fig:overall}
\end{figure}

Our \textbf{major contribution} in this work is to build a
recommendation system for borrowers on \ac{P2PL} platform, which takes
as input the historical loan data with the borrower's characteristic and
outputs the decision on the types of loans they should apply for. Using
our recommendation system, borrowers can achieve a reduced interest rate
payable, with a higher chance of successfully getting the loan request
funded.  The overview of our proposed technique and key technical
contributions are shown in Figure~\ref{fig:overall}.

\begin{itemize}
\item[1] We start with the historical loan dataset. We first filter the
  dataset to remove unusable rows. Next, we encode the categorical
  features to numerical ones.
\item[2] We build machine learning models to predict the interest rate
  payable for bidding and traditional loans along with selecting the
  most important borrower features that influence the models.
\item[3] We build machine learning models to classify if a given
  borrower will succeed on the bidding loan platform along with
  selecting the features that positively influence the model. In steps
  \textcircled{2} and \textcircled{3}, we compare different machine
  learning algorithms, including linear and logistic regression
  (LOGIT)~\cite{neter1996applied}, Random Forest
  (RF)~\cite{feldman2005mortgage}, Support Vector Machine
  (SVM)~\cite{drucker1997support}, and k-nearest neighbour
  (k-NN)~\cite{chatterjee1970nonparametric}.
\item[4] We improve the positive sentiments in the textual description
  for borrowing, which in turn increases chances of bidding loan being
  successfully funded.
\item[5] Using the results from the previous steps, we compare the
  interest rate and success rate of traditional and bidding loans with
  the ideal case: 0\% interest rate and 100\% success rate. The one
  closest to the ideal case would be recommended as the loan type that
  the borrower should apply for.
\end{itemize}

The rest of the paper is organised as
follows. Section~\ref{sec:related-work} reviews and discusses the
current state-of-the-art. Section~\ref{sec:probl-stat-data} describes
the problem statement and the \ac{P2PL}
dataset. Section~\ref{sec:method} describes the details of the workflow
of our proposed technique.  The experimental results and quantitative
comparison with the current state-of-the-art technique is presented in
Section~\ref{sec:result}. Finally, we conclude the paper and discuss the
advantages and limitations of the proposed model in
Section~\ref{sec:concl-future-work}.

\section{Related work}
\label{sec:related-work}

With the burgeoning growth of online \ac{P2PL} marketplaces, a great
deal of research has been proposed to guide lenders and borrowers to
benefit from the \ac{P2PL} system.  From the lender's perspective,
recent work in~\cite{malekipirbazari2015risk} compares different machine
learning algorithms and finds that the best algorithm to predict the
possibility of a loan/borrower defaulting is random forests.  Another
work in~\cite{herzenstein2011strategic} studies the strategic herding
behaviour in \ac{P2PL} loan auctions and points out that the strategic
herding behavior benefits bidders individually and collectively.  Other
works in~\cite{Guo2016417} and~\cite{ren2019investment} proposed
recommendation systems for lenders that yield an investment portfolio
with minimal risk of default along with maximum returns.

From a borrower's perspective, recent works
in~\cite{herzenstein2011tell} and~\cite{althoff2014ask} study the role
of identity claims constructed in narratives by borrowers, and reveals
that as the number of identity claims in narratives increases, the
likelihood of successful funding also increases. Another work
in~\cite{herzenstein2008democratization} studies the determinants of
funding success in online \ac{P2PL} communities and finds out that the
most predominant predictors of loan's likelihood of being funded
successfully are the extent of personal characteristics\footnote{The
  text describing the reason for borrowing money.} provided by
borrowers, and their credit grades. The work in~\cite{ryan2007fund}
weighs the financial and social features of borrowers to determine their
influence in the success of loan being funded. For the purpose of
predicting the likelihood of the loan being funded successfully, the
most recent work in~\cite{ceyhan2011dynamics} explores temporal dynamics
of loan listings and builds a regression model to predict likelihood of
successful funding. However, this model can only yield high accuracy
under the assumption that the bidding process on loans is already
finished. Specifically, the work in~\cite{ceyhan2011dynamics} uses
features called 'number of bids', that are recorded in historical
dataset \textit{only} after the bidding process is completed. We aim to
help \textbf{new} borrowers to make good decisions on the types of loans
to apply for. In turn, this means that features recorded after
completion of bidding cannot be used in our problem setup. In this
paper, we compare different machine learning algorithms with three
feature selection techniques to improve the accuracy of prediction.
Moreover, we also quantitatively compare our technique with the one
proposed in~\cite{ceyhan2011dynamics}.

\section{Problem statement, objectives, and data analysis}
\label{sec:probl-stat-data}

In the next two sections we give our problem statement, objectives, and
describe the historical dataset we have used for analysis and
prediction.

\subsection{The problem statement and objectives}
\label{sec:problem-statement}

In a \ac{P2PL} system, let: \textcircled{1}
$U = \{u_{1}, u_{2}, \dots ,u_{n}\}$ be the set of borrowers,
\textcircled{2} $C = \{c_{1}, c_{2}, \dots, c_{m}\}$ be the set of
features (characteristic of the borrowers) and \textcircled{3}
$V = \{v_{ij}: i \in U, j \in C \}$ be the set of values for each feature of
each borrower, where the entries $v_{ij}$ denote the value of feature
$j$ given by borrower $i$. Furthermore, we use $P_{bid} \subseteq C$ and
$P_{trad} \subseteq C$ as the set of predictors for bidding loans and
traditional loans needed to predict the interest rates $I_{bid}$ and
$I_{trad}$, respectively. In practice, the traditional loans are funded
quickly.  The work in~\cite{barasinska2014crowdfunding} finds the
success rate of getting funded for traditional loans is 81\%. Thus, we
set the success rate of the traditional loan $S_{trad} = 0.81$.
Finally, we use $P_{suc} \subseteq C$ as the set of predictors for success rate
of bidding loans $S_{bid}$.  Overall, we specially focus on the
following research problems:

\begin{itemize}
\item How to accurately predict the interest rate of bidding and
  traditional loans, $I_{bid}$ and $I_{trad}$, respectively.

\item How to accurately predict and increase the success rate of getting
  funded for bidding loans, $S_{bid}$. 
  
  
\item Given the interest rates and success rates of both bidding and
  traditional loans. In order to obtain the lowest interest rate payable
  and increase his/her chances of successfully getting funded, which
  type of loans should the borrower apply for? This can be formulated
  as below:
   $$
  \begin{aligned} 
  &\max        && (-I, S) \\ 
  &\text{ s.t.}&&(I, S) \in \{(I_{trad}, S_{trad}), (I_{bid}, S_{bid})\},\\
  &\text{ and} && I, S \in [0, 1],\\
  \end{aligned}
  $$
  where $I, S \in \mathbb{R}^{\geq0}$.
\end{itemize}

\subsection{Data analysis}
\label{sec:data}

In this paper, we train and test based on dataset from a well
established online \ac{P2PL} platform, Prosper. We choose Prosper,
because Prosper is America's first online \ac{P2PL} platform with over
\$14 billion in funded loans since 2006~\cite{Prop}. In addition,
Prosper has been in operation for more than 10 years, and hence, can
offer a plethora of historical data that is necessary for training and
testing. Other researchers who study \ac{P2PL} systems also use the same
dataset from Propser~\cite{Udacity}, which makes it possible for us to
compare our work with the current state-of-the-art techniques.

In this paper, we use \textit{two} Prosper datasets. \textcircled{1} The
\textit{traditional loan} dataset, which contains 113,938 funded loans
with 81 features in total, dating from 2005 to 2014. \textcircled{2} The
\textit{bidding loan} dataset, which contains 12,774 bidding loans with
30 features in total, dating from 2007-5-27 to 2007-6-30. Not all
features, from both dataset, are applicable to our study, hence we
filter the dataset using the following rules:

\begin{itemize}
\item Remove blank, zero, and missing features, because there is no
  information in these features.
\item Remove features that are \emph{not} applicable to new borrowers,
  for example `number of bids' and `loan current days delinquent'. We
  are aiming to predict the interest rate payable and success rate of
  loan getting funded for \textit{new} borrowers, any feature that are
  recorded after the loan has started cannot be considered.
\item Remove loans with missing values.
\end{itemize}

After filtering both datasets by applying the above rules, we end up
with two datasets that are described below:

\begin{itemize}
\item \textit{Traditional dataset}: contains 70,849 funded loans with 31
  features and 1 response variable --- the borrower's interest rate. Among
  these 31 features, 5 of them are categorical and the rest are
  numerical (see Table~\ref{table:fet}).
\item \textit{Bidding dataset}: contains 12,006 loans (both funded and
  non-funded loans) with 12 features and 2 response variables --- the
  borrower's interest rate and the status of the bidding loan --- funded
  or not funded. Among these 12 features, 6 of them are categorical and
  the rest are numerical (see Table~\ref{table:feb}).
\end{itemize}


\begin{table}[th]
  \centering
  \caption{Features and response variable description of traditional dataset}
  \label{table:fet}
  \resizebox{\columnwidth}{!}{%
    \begin{scriptsize}
      \begin{tabular}{lll}\toprule
        \hline
        Feature   & Explanation    & Type                                                                                                                                                                              \\ \midrule
        BorrowerRate                       & The Borrower's interest rate for this loan.                              & Numerical                                                                                                                    \\ 
        OpenCreditLines                    & Number of open credit lines.                             & Numerical                                                                                                                                    \\  
        ProsperGrade            & A custom rating score built by Prosper.                 &      Categorical                                                                                                             \\ 
        ProsperScore                       & A custom risk score built by Prosper.                            & Numerical                                                                                                         \\ 
        ListingCategory         & The category of the listing that the borrower.                      & Numerical                                                                                     \\
        CurrentCreditLines                 & Number of current credit lines.                               & Numerical                                                                                                                               \\ 
        TotalCreditLinespast7years         & Number of credit lines in the past seven years.                             & Numerical                                                                                                                 \\ 
        OpenRevolvingAccounts              & Number of open revolving accounts.                            & Numerical                                                                                                                               \\ 
        OpenRevolvingMonthlyPayment        & Monthly payment on revolving accounts.                             & Numerical                                                                                                                          \\ 
        TotalInquiries                     & Total number of inquiries.                            & Numerical                                                                                                                                       \\ 
        CurrentDelinquencies               & Number of accounts delinquent.                           & Numerical                                                                                                                                    \\ 
        AmountDelinquent                   & Dollars delinquent.                          & Numerical                                                                                                                                                \\ 
        Occupation                         & The Occupation selected by the Borrower.                           & Categorical                                                                                                                          \\
        PublicRecordsLast10Years           & Number of public records in the past 10 years.                               & Numerical                                                                                                                \\ 
        RevolvingCreditBalance             & Dollars of revolving credit.                             & Numerical                                                                                                                                    \\ 
        TradesNeverDelinquent & Trades that have never been delinquent.                          & Numerical                                                                                                              \\ 
        TotalTrades                        & Number of trade lines ever opened.                              & Numerical                                                                                                                             \\ 
        StatedMonthlyIncome                & The monthly income the borrower stated.                              & Numerical                                                                                                                        \\ 
        AvailableBankcardCredit            & The total available credit via bank card.                                 & Numerical                                                                                                                   \\ 
        TradesOpenedLast6Months            & Number of trades opened in the last 6 months.                              & Numerical                                                                                                                  \\ 
        BankcardUtilization                & The percentage of available credit that is utilized.                            & Numerical                                                                                                   \\ 
        Homeownership                & Specifies if the borrower is a homeowner or not.                                & Categorical                                                                                                        \\ 
        DebtToIncomeRatio                  & The debt to income ratio of the borrower.                            & Numerical                                                                                                                        \\ 
        InquiriesLast6Months               & Number of inquiries in the past six months.                              & Numerical                                                                                                                    \\ 
        LoanAmount                 & The origination amount of the loan.                                & Numerical                                                                                                                          \\ 
        CreditScoreRangeLower              & The lower range of the borrower's credit score.                                & Numerical                                                                                       \\ 
        EmploymentStatusDuration           & The length in months of the employment status.                             & Numerical                                                                                                                  \\ 
        DelinquenciesLast7Years            & Number of delinquencies in the past 7 years.                             & Numerical                                                                                                                    \\ 
        Term                               & The length of the loan expressed in months.                            & Numerical                                                                                                                      \\ 
        BorrowerState                      & The state of the address of the borrower.                           & Categorical                                                                                          \\ 
        EmploymentStatus                   & The employment status of the borrower.                          & Numerical                                                                                                                            \\ 
        Description                         & The description of the lowan written by the borrower.                                & Categorical                                                                                                                                                                                            \\  \bottomrule
      \end{tabular}%
    \end{scriptsize}
  }
\end{table}

\begin{table}[th]
  \centering
  \caption{Features and response variable description of bidding dataset}
  \label{table:feb}
  \resizebox{\columnwidth}{!}{%
    \begin{scriptsize}
      \begin{tabular}{lll}\toprule
        \hline
        Feature   & Explanation    & Type                                                                                                                                                                              \\ \midrule
        BorrowerRate                       & The Borrower's interest rate for this loan.                              & Numerical                                                                                                                    \\ 
        BorrowerMaximumRate                    & The maximum interest rate the borrower will accept.                             & Numerical                                                                                                                                    \\  
        ProsperGrade            & A custom rating score built by Prosper.                 &      Categorical                                                                                                             \\ 
        Homeownership                & Specifies if the borrower is a homeowner or not.                                & Categorical                                                                                                        \\ 
        DebtToIncomeRatio                  & The debt to income ratio of the borrower.                            & Numerical                                                                                                                        \\ 
        LoanAmount                 & The origination amount of the loan.                                & Numerical                                                                                                                          \\ 
        FundingOption          & The options of funding.                             & Categorical                                                                                                                  \\ 
        Images            & Number of images that are uploaded by borrowers.                             & Numerical                                                                                                                    \\ 
        Duration                               & The length of funding duration.                            & Numerical                                                                                                                      \\ 
        BorrowerState                      & The state of the address of the borrower.                           & Categorical                                                                                          \\ 
        EmploymentStatus                   & The employment status of the borrower.                          & Numerical                                                                                                                            \\ 
        HasVerifiedBankAccount                   & Specifies if or not the bank account is verified.                                & Categorical                                                                                                       \\ 
        Description                         & The description of the lowan written by the borrower.                                & Categorical                                                                                                 \\
        LoanStatus                         & The current status of the loan.                                & Categorical                                                                                      \\  \bottomrule
      \end{tabular}%
    \end{scriptsize}
  }
\end{table}

\section{Methodology}
\label{sec:method}

In this section, we first introduce the method we use to encode
categorical features followed by sentiment analysis for textual
descriptions of reasons for borrowing, as input by the borrowers. Next,
we introduce the feature selection techniques and machine learning
algorithms we use to predict interest rate payable for bidding and
traditional loans, respectively, along with computing the chances of
success of any given bidding loan (recall that the chance of success of
traditional loan is fixed at $81\%$). Finally, we propose the method to
advice borrowers with the type of loans they should apply for.

\subsection{Feature encoding and sentiment analysis}
\label{sec:feat-encod-sent}

Both bidding and traditional dataset have several categorical
features. These features need to be transferred to a numerical value, so
that they can be used in machine learning algorithms like linear
regression, \ac{LOGIT}, etc. In the biding and traditional datasets, we
split the categorical features into three types:

\begin{itemize}
\item[1] Features that only have two classes such as ``Homeownership''
  and ``Funding option''.
\item[2] Features that have more than two classes like ``Prosper grade''
  and ``Borrower's state''.
\item[3] Features that contain random textual (English) words like
  Borrower's ``Description'' of the loan.
\end{itemize}

\begin{table}[t]
  \begin{center}\caption{A simple example: characteristic of two borrowers}\label{Tab:ID}
    \resizebox{\columnwidth}{!}{%
      \begin{tabular}{llll} \toprule
        \hline 
        Feature                       & Borrower 1            & Borrower 2                     \\ \midrule
        Borrower maximum rate         & 0.16                  & 0.105                            \\ 
        Prosper grade                 & 7 (HR)                & 1 (AA)                           \\ 
        Term                          & 36                    & 36                                 \\ 
        Credit score                  & 540                   & 760                                 \\ 
        Delinquencies in last 7 years & 5                     & 0                                     \\ 
        Debt to income ratio          & 0.17                  & 0.06                                 \\ 
        Loan amount                   & 2,300                 & 10,000                              \\ 
        Homeownership                 & 0 (Not own)           & 1 (Own)                            \\ 
        Duration                      & 3                     & 10                                 \\ 
        Funding option                & 0 (Close when funded) & 1 (Open for duration)  \\ 
        Has verified bank account     & 1 (True)              & 1 (True)                          \\ 
        Images                        & 0                     & 0                                  \\ 
        Description                   & 0.3818 (Payoff Credit Cards)   & 0 (Lender seeing Prosper from borrower's point-of-view) \\
        \bottomrule
      \end{tabular}%
    }
  \end{center}
\end{table}

Table~\ref{Tab:ID} details the features of interest for two borrowers.
In order to encode the categorical features, we use two most popular
encoding techniques: \textit{binary encoding} and \textit{ordinal
  encoding}. Binary encoding technique transfers each categorical
feature into \textit{new} numerical features containing only zeros and
ones. For instance, the categorical feature ``Homeownership'' has two
classes ``Own'' or ``Not own''. Each of these classes is encoded as an
individual numerical feature, with a value of 0 or 1.  In case of the
first borrower in Table~\ref{Tab:ID}, the ``Homeownership'' feature will
be translated into two new features ``Not own:1'' and ``Own:0''. However,
either of these two new features is enough to show the home ownership 
of the borrower. To avoid the increase of the number of features, we only
select one of them as the encoded feature. In other words, the feature 
``Homeownership'' is encoded with a value of 0 (Not own) or 1 (Own). Same
for borrower two.

Binary encoding is only feasible for categorical features with few
classes. In case of a categorical feature with a plethora of classes,
this method will increase the total number of features in the dataset
significantly. Ordinal encoding is the preferred method in such cases.
It converts string labels to integer values 1 through $k$, where $k$ is
the number of classes in a given categorical feature. For example,
consider the feature ``Proser grade'' shown in Table~\ref{Tab:ID}. There
are 7 classes in this feature: AA, A, B, C, D, E, HR.\@ After applying
ordinal encoding, the 7 classes are transferred to 1, 2, 3, 4, 5, 6 and
7, respectively. Finally, neither binary encoding nor ordinal encoding
is applicable when encoding the third type of categorical feature:
textual data, because textual description has meaning, which should be
captured by the encoding technique.

In order to encode the third type of feature, we do sentiment analysis.
Sentiment analysis can extract and evaluate the emotions in text. There
are two popular types of sentiment analysis: \textcircled{1} classify
the polarity of given text as positive, negative or neutral.
\textcircled{2} Evaluate a given piece of text to a certain score, which
shows the levels of emotion. In this paper, we apply the second type of
sentiment analysis to encode the text as numerical value. Specifically,
we apply the sentiment analysis technique from VADER (Valence Aware
Dictionary and sEntiment Reasoner)~\cite{hutto2014vader}, thereby
encoding the text into numerical scores, which we call the
\textit{sentiment score}, ranging from -1 to 1. Here, 1 represents the
most positive emotion and -1 repents the most negative emotion. For
example, in Table~\ref{Tab:ID}, the description ``Payoff Credit Cards''
(as the reason for borrowing money) is evaluated to 0.3818. An optimal
sentiment score can help with getting the loan funded. The results
comparing the likelihood of getting funded with varying sentiment scores
are described in Section~\ref{sec:senti}.

\subsection{Machine learning models for interest rate prediction,
  likelihood of getting funded, and feature selection}
\label{sec:mach-learn-models}

Recall that there are 31 features in the traditional dataset and 12
features in the bidding dataset (Section~\ref{sec:data}). However, not
all of these features are useful when predicting the interest rate
and/or the success rate of getting funded. Moreover, the machine
learning algorithms we use in this paper such as \acs{SVM} and
\acs{k-NN} are sensitive to irrelevant features. Hence, it is necessary
to do feature selection along with predicting the interest rates payable
and the likelihood of getting funded.

In this paper, we compare three popular feature selection algorithms:
\textcircled{1} forward selection, \textcircled{2} backward selection
and \textcircled{3} recursive selection. Since all the three techniques
are well known, here we only give an overview of the techniques and the
results are shown in Sections~\ref{sec:PIR} and~\ref{sec:PLS}. In
forward selection, we start with an empty feature set ($P_{trad}$,
$P_{bid}$, and $P_{suc}$) and keep on adding new features one after
another until the coefficient of determination/recall rate stops
increasing. In backward selection, we start all the features in the
feature sets, and keep on dropping features one after another until the
coefficient of determination/recall rate stops increasing. In recursive
selection, we start with a set of all the features in the feature sets
and keep on dropping the least important features\footnote{Least
  important features are obtained after fitting the model using the
  scikit learn Python library.} until the coefficient of
determination/accuracy stops increasing.

In order to predict the interest rates of bidding and traditional loans,
we choose four regression models to compare and select the best fitted
model. These regression models include linear regression, \ac{RF},
\ac{SVM} and \ac{k-NN}. All four models have both advantages and
disadvantages, and we apply feature selection on each of the model to
find the best fit. Different from predicting the interest rates, we
select four machine learning classifiers to predict the likelihood of a
bidding loan getting funded. \ac{RF}, \ac{SVM}, and \ac{k-NN} are all
applicable for classification and regression problem, we only replace
linear regression with \ac{LOGIT} to be the fourth classifier. Again, we
apply feature selection on each of the classifier and find the
classifier that results in the highest accuracy (recall rate using cross
validation techniques). The results are described in
Section~\ref{sec:result}.

\subsection{The decision process to recommend the type of loan
  application}
\label{sec:decis-proc-recomm}

Our final goal is to help new borrowers decide, which type of loan they
should apply for. The goal is to achieve the highest likelihood of
successfully getting funded at the lowest interest rate payable. To
reach this goal, we first compute the interest rate payable and the
likelihood of success using the models described in
Section~\ref{sec:mach-learn-models}. The machine learning models output
two tuples: $(I_{trad}, 0.81)$ and $(I_{bid}, S_{bid})$. We next compare
these two tuples with the ideal case: $(0, 1)$, where the first element
of the tuple indicates 0\% interest rate payable and the second
indicates 100\% likelihood of successfully getting funded. The final
decision is made by comparing the Euclidean distance between each of the
tuples obtained from the machine learning algorithms and the ideal case.
We can formalise the approach as follows:

\begin{equation}\label{eq:optm}
  \begin{aligned} 
    &\min        && |(I,S) - (I_{ideal}, S_{ideal})| \\ 
    &\text{ s.t.}&&  (I_{ideal}, S_{ideal}) =  (0, 1),\\ 
    &            &&(I, S) \in \{(I_{trad}, S_{trad}), (I_{bid}, S_{bid})\},\\
    &\text{ and}  && I, S \in [0, 1],\\
  \end{aligned}
\end{equation} 

\noindent
where $|\cdot|$ represents the Euclidean distance, and $I, S \in \mathbb{R}^{\geq0}$.

\section{Experimental results}
\label{sec:result}

In this section, a thorough comparison of the various techniques
described in Section~\ref{sec:method} is presented. 

\subsection{Experimental setup}
\label{sec:ES}

All our experiments are performed on the traditional and bidding Prosper
datasets obtained from~\cite{Udacity} and cleaned/analysed as described
in Section~\ref{sec:data}. The detail of datasets and experiments
performed is presented below:

\begin{itemize}
\item To predict the interest rate of traditional loans we randomly
  sample 10,000 loans from the 70,849 loans available to us. For
  training and testing, we perform 5-fold Montecarlo cross validation by
  splitting the 10,000 loans into a ratio of 80:20. The results of the 5
  runs and the average are shown in Section~\ref{sec:PIR}.
\item To predict the interest rate of bidding loans we sample 908 funded
  bidding loans from the 12,006 available in the bidding dataset. For
  training and testing, we perform 5-fold Montecarlo cross validation by
  splitting the 908 loans into a ratio of 80:20 for training and
  testing, respectively. The results of the 5 separate runs and the
  average results are shown in Section~\ref{sec:PIR}.
\item To predict the success rate of bidding loans we sample 908 funded
  and 908 non-funded loans from the bidding dataset to get in total 1816
  loans. Again, for training and testing, we perform 5 fold Montecarlo
  cross validation, with a 80:20 split for each run, the results are
  presented in Section~\ref{sec:PLS}.
\item The results of sentiment scores impacting success rate of getting
  funded and the overall efficacy of the recommendation system are
  presented in Sections~\ref{sec:senti} and~\ref{sec:comp-with-hist}.
\end{itemize}

\subsection{Interest rate payable prediction for traditional and bidding
  loans}
\label{sec:PIR}

\begin{figure}[!t]
  \centering
  \includegraphics[width=1.0\linewidth]{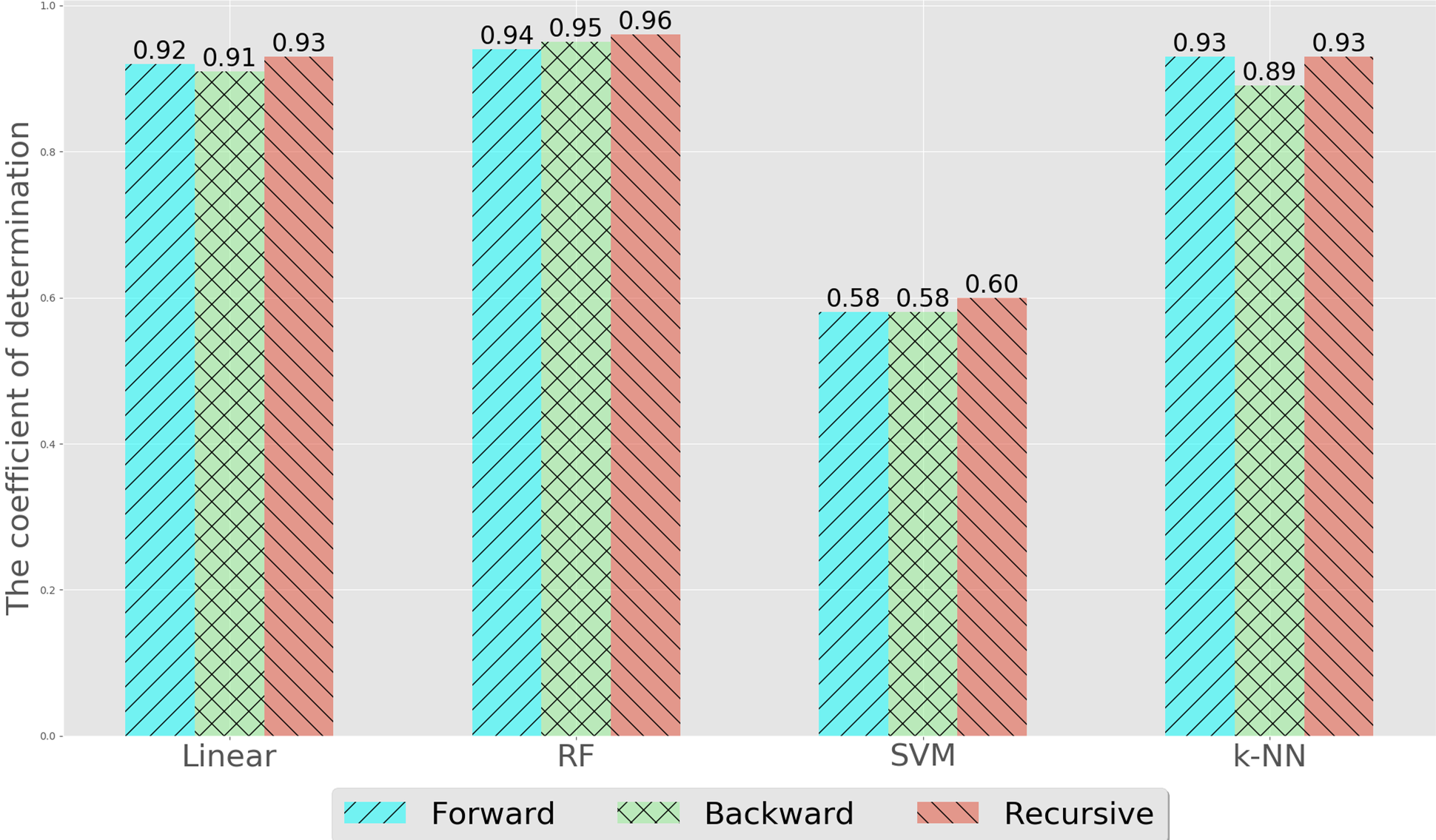}
  \caption{Prediction of interest rates for traditional loans.}
  \label{fig:Itrad}
\end{figure}

In this section, we compare four machine learning regression models with
feature selection techniques to predict the interest rates payable for
both traditional and bidding loans. Since the response variable,
interest rate, is continuous, the coefficient of determination $R^{2}$
is used to evaluate the accuracy of models for comparison purposes. The
coefficient of determination can be formulated as:

\[
  R^{2} = \frac{\text{Residual sum of squares}}{\text{Total sum of
      squares}}
\]

\noindent
where $R^{2}$ ranges from 0 to 1. The larger the value of $R^{2}$, the
better the goodness of fit of a model. Specifically, an $R^{2}$ value
equal to 1 means the regression model perfectly fits the data.

\begin{table}[t]
  \caption{The 5 Monte-Carlo CV of the four regression models with
    recursive selection for predicting the interest rate of traditional
    loans}\label{tab:mcr}
  \begin{tabular}{lllllll} \toprule
    CV test & 1    & 2    & 3    & 4    & 5    & Average \\ \midrule
    Linear  & 0.91 & 0.94 & 0.94 & 0.93 & 0.92 & 0.93    \\
    RF      & 0.93 & 0.97 & 0.96 & 0.97 & 0.96 & 0.96    \\
    SVM     & 0.57 & 0.62 & 0.60 & 0.59 & 0.61 & 0.60    \\
    k-NN    & 0.90 & 0.94 & 0.94 & 0.93 & 0.95 & 0.93   \\ \bottomrule
  \end{tabular}
\end{table}

Figure~\ref{fig:Itrad} illustrates the average performance of predicting
the interest rates payable, for 5 cross validation runs, for traditional
loans by applying linear regression, \ac{RF}, \ac{SVM} and \ac{k-NN}
with forward, backward and recursive feature selection. From
Figure~\ref{fig:Itrad}, we can observe that under the same feature
selection technique, \ac{RF} always performs the best. This result also
can be seen from Table~\ref{tab:mcr}, which shows the 5 Monte-Carlo
cross validation runs for each model with recursive feature selection.
Recursive feature selection outperforms forward and backward feature
selection under the same regression model. In addition, \ac{RF} with
recursive selection achieves the highest coefficient of determination
with value 0.96. Linear regression and \ac{k-NN} also fit the data well
with recursive selection with $R^{2} = 0.93$. However, \ac{SVM} does not
perform well with the value of $R^{2}$ of only 0.6. We choose \ac{RF}
with recursive selection to be the best method to predict the interest
rate of traditional loans. The features selected by recursive feature
selection technique with \ac{RF} are: Prosper grade, term, credit score
and delinquencies in last 7 years.

\begin{figure}[!t]
  \centering
  \includegraphics[width=1.0\linewidth]{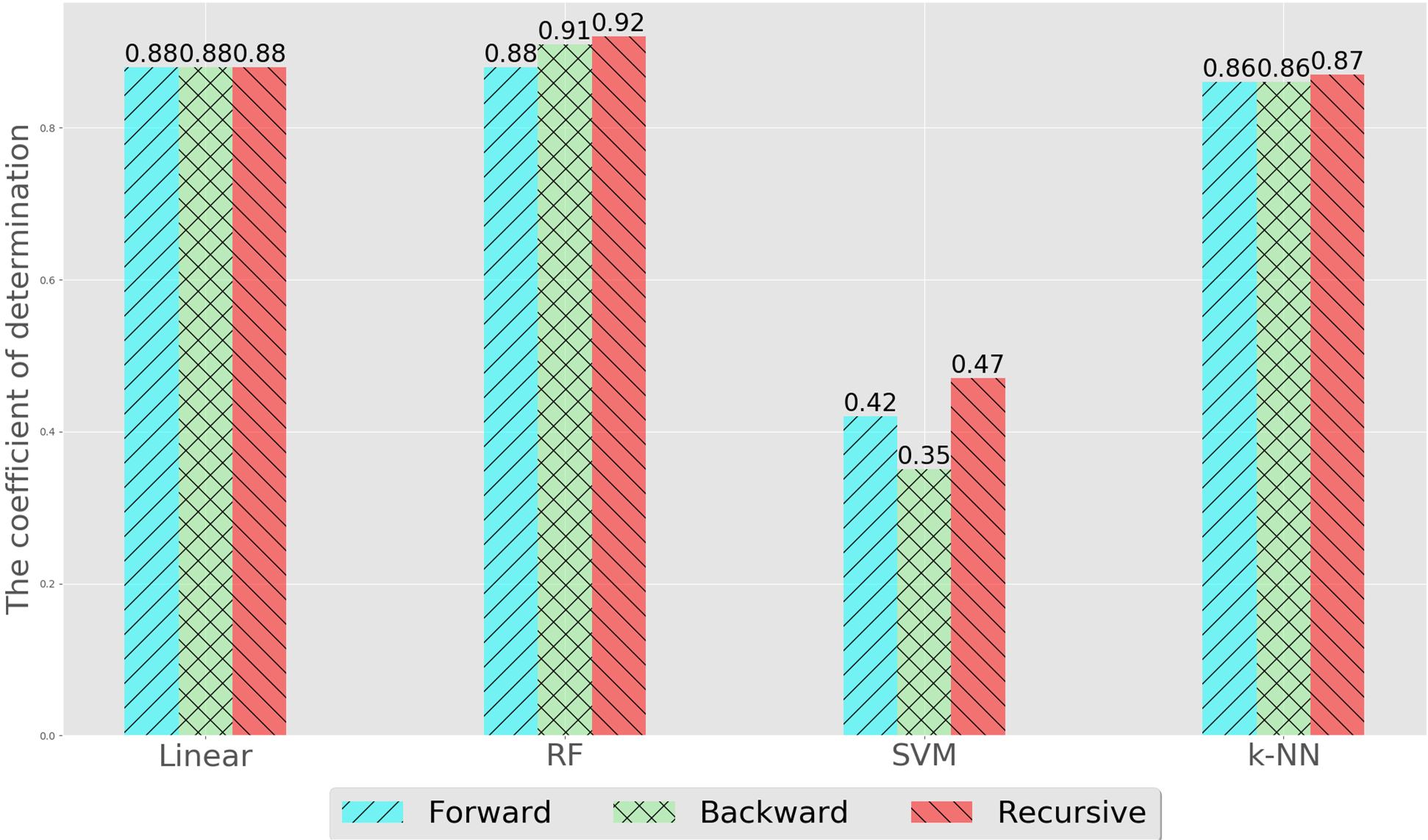}
  \caption{Prediction of interest rates for bidding loans.}
  \label{fig:Ibid}
\end{figure}

\begin{table}[t]
  \caption{The 5 Monte-Carlo CV of the four regression models with
    recursive selection for predicting the interest rate of bidding
    loans}\label{tab:mcb}
  \begin{tabular}{lllllll} \toprule
    CV test & 1    & 2    & 3    & 4    & 5    & Average \\ \midrule
    Linear  & 0.89 & 0.88 & 0.91 & 0.87 & 0.86 & 0.88    \\
    RF      & 0.92 & 0.93 & 0.92 & 0.93 & 0.90 & 0.92    \\
    SVM     & 0.47 & 0.45 & 0.46 & 0.49 & 0.48 & 0.47    \\
    k-NN    & 0.89 & 0.87 & 0.89 & 0.85 & 0.84 & 0.87   \\ \bottomrule
  \end{tabular}
\end{table} 

Next, we predict the interest rates of bidding loans.
Figure~\ref{fig:Ibid} illustrates the average performance of predicting
the interest rates payable, for 5 cross validation runs, for bidding
loans by applying linear regression, \ac{RF}, \ac{SVM} and \ac{k-NN}
with forward, backward and recursive feature selection. It can be
observed from Figure~\ref{fig:Ibid} that under the same feature
selection technique, \ac{RF} performs best. The highest $R^{2}$ of 0.92
is achieved by applying \ac{RF} with recursive feature selection. Linear
regression and \ac{k-NN} also perform well with the values of $R^{2}$ of
0.88 and 0.87, respectively. In addition, with the same regression
model, recursive feature selection gives the best subset of features
that results in the highest value of $R^{2}$. This result also can be
observed from Table~\ref{tab:mcb}, which describes the 5 Montecarlo
cross validation runs with recursive feature selection. \ac{SVM} again
does not perform well when predicting the interest rate payable for
bidding loans with the value of $R^{2}$ of only 0.47. Therefore, for the
purpose of accurately predicting the interest rate of bidding loans, we
select \ac{RF} with recursive feature selection as the preferred
prediction model. The selected features are: borrower maximum rate,
Prosper grade, debt to income ratio, loan amount, homeownership,
duration, funding option and has verified bank account.

\subsection{Predicting the success rate of funding bidding loans}
\label{sec:PLS}

In this section, we compare \ac{LOGIT}, \ac{RF}, \ac{SVM}, and \ac{k-NN}
with forward, backward and recursive feature selection techniques to
find the best classification model for predicting the success rates of
getting funded for bidding loans. Since the response variable here is
either funded or non-funded, we select the accuracy (recall rate) as the
criterion to evaluate the goodness of fit of the model. The accuracy
measure used for comparison purpose can be formulated as follows:

\[
  \text{Accuracy } = \frac{\text{Number of correct
      predictions}}{\text{Total number of predictions}}
\]

\noindent
where the accuracy $\in [0, 1]$. The higher the accuracy, the better the
model fits the dataset.

\begin{figure}[!t]
  \centering
  \includegraphics[width=1.0\linewidth]{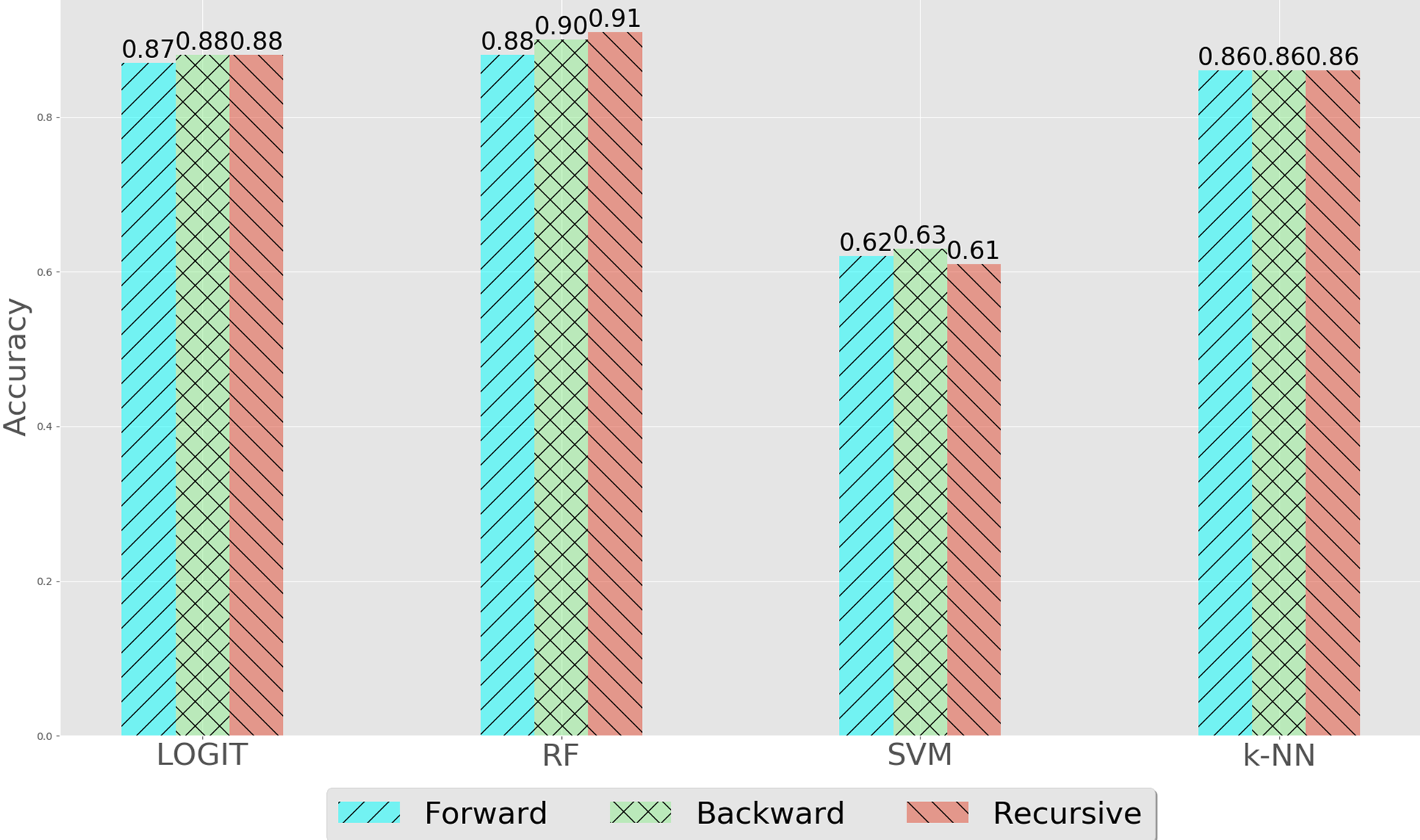}
  \caption{Prediction of the likelihood of successfully getting funded
    for bidding loans.}
  \label{fig:Sbid}
\end{figure}

\begin{table}[t]
  \caption{The 5 Monte-Carlo CV of the four regression models with
    recursive selection for predicting the success of getting funded of
    bidding loans}\label{tab:mcs}
  \begin{tabular}{lllllll} \toprule
    CV test & 1    & 2    & 3    & 4    & 5    & Average \\ \midrule
    LOGIT   & 0.86 & 0.87 & 0.89 & 0.88 & 0.90 & 0.88    \\
    RF      & 0.89 & 0.93 & 0.90 & 0.92 & 0.92 & 0.91    \\
    SVM     & 0.59 & 0.61 & 0.63 & 0.60 & 0.61 & 0.61    \\
    k-NN    & 0.87 & 0.85 & 0.89 & 0.87 & 0.88 & 0.87   \\ \bottomrule
  \end{tabular}
\end{table}

Figure~\ref{fig:Sbid} shows the average accuracy, from amongst the 5
cross validation runs, of predicting the success of getting funded for
bidding loans by applying \ac{LOGIT}, \ac{RF}, \ac{SVM} and \ac{k-NN}
with forward, backward and recursive feature selection. The results of
the 5 Monte-Carlo cross validation runs with recursive feature selection
are shown in Table~\ref{tab:mcs}. From Figure~\ref{fig:Sbid}, it can be
seen that with the same feature selection technique, \ac{RF} has the
best average accuracy. In addition, the performance of \ac{SVM} is bad
around 62\% accuracy. Therefore, to predict the success of getting
funded of bidding loans as accurately as possible, we select the \ac{RF}
together with recursive feature selection as the preferred prediction
model.

\begin{table}[t]
  \caption{The confusion matrix of \ac{RF} with recursive selection on
    test bidding dataset}\label{tab:cm}
  \begin{tabular}{lll}\toprule
    & Actual funded      & Actual non-funded  \\ \midrule
    Predicted funded     & 161 true positives & 20 false positives \\
    Predicted non-funded & 12 false negatives & 171 true negatives \\ \bottomrule
  \end{tabular}
\end{table}

Table~\ref{tab:cm} shows the confusion matrix of \ac{RF} with recursive
feature selection for a single Montecarlo run . We can observe that the
false positives and false negatives
are small compared to the true positives and the true negatives.
Specifically, the true positive rate is 93\% and the true negative rate
is 90\%. These results validate that our proposed model is reliable when
predicting the success of getting funded for bidding loans.

In order to verify that the proposed technique is better than the
current state-of-art model~\cite{ceyhan2011dynamics}, we randomly split
the bidding dataset 1816 loans (908 funded and non-funded loans) to
training and testing dataset with ratio of 80:20. Then we train the
\ac{RF} model with recursive feature selection and the current
state-of-the-art model on the training dataset (1452 loans) and test on
the testing dataset (364 loans). The results are shown in
Table~\ref{Tab:preds}. We can observe from Table~\ref{Tab:preds} that
our proposed model has 0.91 accuracy, which is 0.24 (24\%) higher than
the current state-of-the-art technique. Hence, we can state that the
\ac{RF} technique with the recursive feature selection algorithm
outperforms the current state-of-the-art technique (as proposed
in~\cite{ceyhan2011dynamics}).

%

\begin{table}[!t]
  \begin{center}\caption{The accuracy of predicting the success of getting funded of bidding loans}\label{Tab:preds}
    \begin{tabularx}{\columnwidth}{p{1.34cm}lp{4.5cm}}\toprule
      Algorithm & Accuracy & Selected features \\ \midrule
      RF with recursive selection              & \textbf{0.91}     & Borrower maximum rate, Prosper grade, debt to income ratio, loan amount, homeownership, funding option, has verified bank account, images, sentiment score \\
      Current state-of-the-art & 0.67 & Borrower maximum rate, debt to
      income ratio, loan amount, homeownership, listing description
      length \\ \bottomrule
    \end{tabularx}
  \end{center}
\end{table}

\subsection{Impact of sentiment score}
\label{sec:senti}

\begin{figure}[!t]
  \centering
  \includegraphics[width=1.0\linewidth]{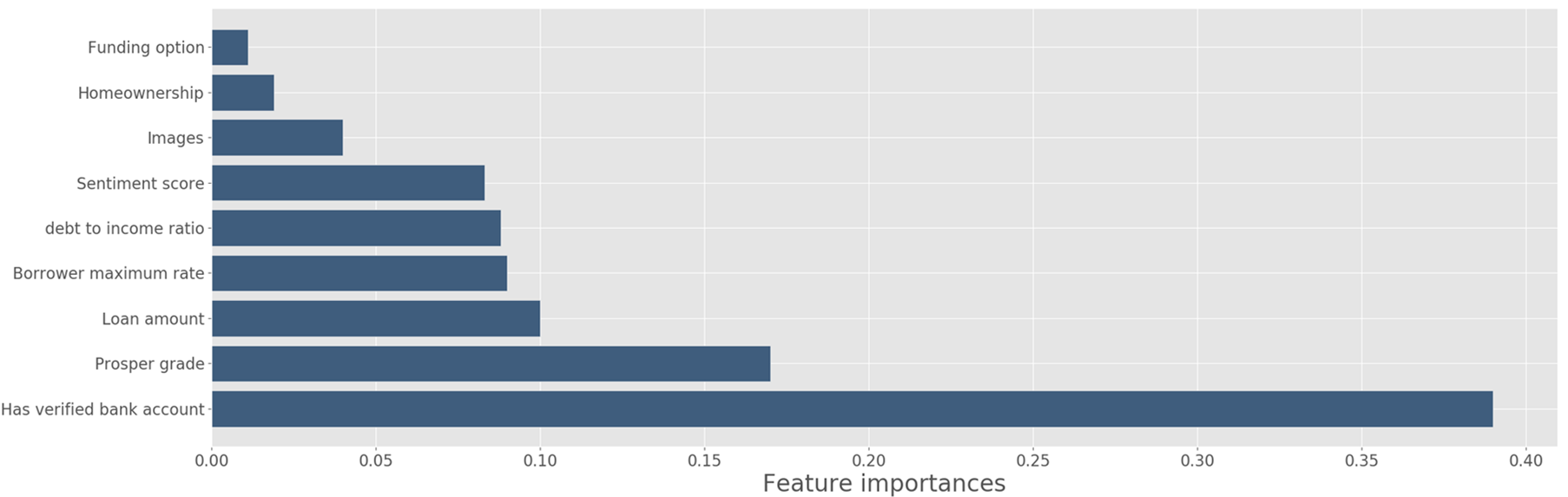}
  \caption{Feature importances of RF with recursive feature selection
    model}
  \label{fig:feat}
\end{figure}

In this section we apply the sentiment analysis technique from
VADER~\cite{hutto2014vader} on the bidding dataset and study the impact
of raising the sentiment score on the likelihood of getting funded for
bidding loans. Recall that \ac{RF} with recursive feature selection is
the best model to predict the success rate of getting funded for bidding
loans. The feature ``sentiment score'' is selected in the predictors
(see Table~\ref{Tab:preds}). Recall that the feature sentiment score is
encoded by applying sentiment analysis from VADER on the textual
``description'' for borrowing. The selection of sentiment score, as a
predictor, indicates that the emotion of texts written by borrowers does
impact the success rate of getting funded when applying for bidding
loans. Figure~\ref{fig:feat} gives the feature importances of the nine
features selected by \ac{RF} with recursive feature selection technique.
From Figure~\ref{fig:feat}, we can observe that sentiment score is the
$6^{th}$ most influential feature.

\begin{figure}[!t]
  \centering
  \includegraphics[width=1.0\linewidth]{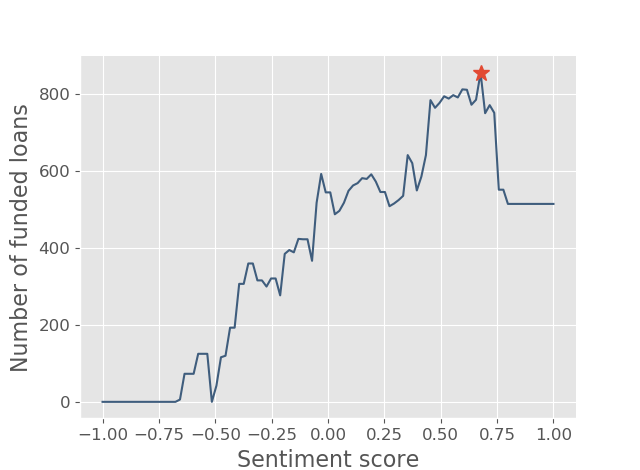}
  \caption{Impact on success of getting funded by changing the sentiment
    of texts.}
  \label{fig:senti}
\end{figure}

We want to observe if crafting the ``description'' for borrowing can
help improve the overall success rate of getting funded. In order to do
so, we control the sentiment score of all 11,098 non-funded loans, from
the bidding dataset, from -1 to 1 and then apply the proposed success
rate prediction model to see how many of them will become funded.
Figure~\ref{fig:senti} shows the number of funded loans by changing the
sentiment score. We can observe that by changing the sentiment score to
anywhere between $(0.45, 0.70)$, around 800 non-funded loans are
transferred to funded. In addition, when sentiment score equals 0.68,
the maximum number of loans (856) get funded. These results indicate
that borrower' should write the description with a more positive
sentiment. However if the description is \textit{too} positive, the
chance of getting funded decreases, which can be seen from
Figure~\ref{fig:senti}. This is because too much positive sentiment
looks fake.

\begin{table}[!t]
  \begin{center} \caption{Results of increasing sentiment score on the
      12,006 loans in the bidding dataset.} \label{tab:com}
    \resizebox{\columnwidth}{!}{%
      \begin{tabular}{lll} \toprule
        Dataset                          & Funded loan & Non-funded loan\\ \midrule
        No change in sentiment score   & 908             & 11098                 \\
        Increasing the sentiment score to a positive value & 1764           & 10242                \\ \bottomrule
      \end{tabular}%
    }
  \end{center}
\end{table}

Table~\ref{tab:com} compares the number of originally funded loans in
the bidding dataset and the new dataset obtained by changing the
sentiment score to a positive value. It can be observed that the number
of funded loans in the resultant dataset, which is obtained by
increasing the sentiment score, is about twice that in the original
dataset. Performing a t-test with the null hypothesis that these two
dataset have the same mean, the p-value is just 1.67e-69. Hence, we can
deduce that our proposed method can potentially improve the chance of
success of getting funded significantly.

\subsection{Efficacy of the overall recommendation engine}
\label{sec:comp-with-hist}

In this section we present the results of the overall efficacy of the
proposed recommendation engine. We first sample 1000 loans from the
historical dataset; 500 from the traditional and 500 from the bidding
dataset, respectively. Next, we apply \ac{RF} with recursive feature
selection to obtain the interest rates payable ($I_{trad}$, $I_{bid}$)
for each loan. Next, we compute the success rate of the loans
($S_{bid}$) after setting the sentiment score to 0.68 (the optimal
value), recall that $S_{trad} = 0.81$. Finally, we apply
Equation~(\ref{eq:optm}) to recommend to the borrower if they should be
applying for traditional or bidding loan.

The results are shown in Table~\ref{tab:cphis}. From
Table~\ref{tab:cphis}, we can observe that most of the borrowers who
apply for the bidding loans are recommended to apply for traditional
loans. This results is expected, because most biding loans, even after
increasing the sentiment scores, remain unfunded. However, the number of
funded loans increases from 569 to 820, and the average interest rate of
funded loans is decreased by 3\%. These results show that our proposed
recommendation system can help borrowers to get fund successfully with a
lower interest rate.

\begin{table}[t]
  \caption{Comparison between  the historical data and the results obtained by applying the proposed mehtod}
  \label{tab:cphis}
  \resizebox{\columnwidth}{!}{%
    \begin{tabular}{llllll} \toprule
      & Traditional loans & Bidding loans & Funded & Non-funded & Average interest rate of funded loans   \\ \midrule
      Historical/original dataset                                     & 500     & 500    & 569        & 431                                   & 0.23 \\
      The proposed method & 892     & 108    & 820        & 180                                   & 0.20 \\ \bottomrule
    \end{tabular}%
  }
\end{table}

\section{Conclusion and future work}
\label{sec:concl-future-work}

Online \acf{P2PL} marketplaces, connect lenders directly to borrowers.
The convince of use has led to a burgeoning growth of \ac{P2PL}
marketplaces. Borrowers using \ac{P2PL} marketplaces are usually looking
to get a loan at the lowest interest rate. \ac{P2PL} marketplaces
provide two main pathways for obtaining loans: \textcircled{1}
traditional loans, where the platform decides the interest rate for the
borrowers and the lenders fund the loan, or \textcircled{2} the lenders
themselves decide on the interest they want, by bidding on the loan.
Borrowers can get a lowered interest rate via the bidding technique,
however with a reduced likelihood of getting funded. Hence, it is
essential to make individual recommendation to borrowers depending upon
their situation. However, to the best of our knowledge no such
recommendation system exists.

We build a recommendation system for borrowers that recommends the best
loan option for them, which results in higher likelihood of getting
funded, while reducing the interest rate payable. Our methodology
consists of three main steps. In step-\textcircled{1}, use \ac{RF} with
recursive borrower feature selection model to predict the interest rates
of bidding and traditional loans. In step-\textcircled{2}, we use
\ac{RF} with recursive borrower feature selection model to predict the
success rate of getting funded for bidding loans and even improve the
sentiment of reasons for borrowing. Finally, in step-\textcircled{3},
given the interest rates and success rates of both traditional and
bidding loans, we compare them with the ideal case and determine the
best choice for borrowers.

Experimental results show that our proposed method outperforms the
current state-of-the-art technique, in that the accuracy of correctly
predicting the success rate of bidding loans increases from 67\% to
91\%. In addition, the proposed technique can increase the chances of
getting funded by $2\times$. The main drawback of our proposed method is that
it currently cannot craft the reason for borrowing in order to increase
the sentiment scores, we plan to remedy this in the future.

\bibliographystyle{ACM-Reference-Format}
\bibliography{mybibfile}

\end{document}

%% file: 00_acronyms.tex
\acrodef{P2PL}{Peer to Peer Lending}
\acrodef{ARR}{Annualized Rate of Return}
\acrodef{LOGIT}{Logistic Regression}
\acrodef{MLOGIT}{Multi-nominal Logistic Regression}
\acrodef{SVM}{Support Vector Machine}
\acrodef{RF}{Random Forest}
\acrodef{FICO}{Fair, Isaac, and Company}
\acrodef{H3LC}{Hierarchical Three Label Classification}
\acrodef{MSE}{Mean Square Error}
\acrodef{RMSE}{Root Mean Square Error}
\acrodef{RBF}{Gaussian Radial Basis Kernel Function}
\acrodef{EDM}{Exponential Distance Method}
\acrodef{CV}{Cross Validation}
\acrodef{NNs}{Neural Networks}
\acrodef{GP}{Gaussian Process}
\acrodef{k-NN}{$k$-Nearest Neighbors}

%% file: conference.bbl

\begin{thebibliography}{17}


\ifx \showCODEN    \undefined \def \showCODEN     #1{\unskip}     \fi
\ifx \showDOI      \undefined \def \showDOI       #1{#1}\fi
\ifx \showISBNx    \undefined \def \showISBNx     #1{\unskip}     \fi
\ifx \showISBNxiii \undefined \def \showISBNxiii  #1{\unskip}     \fi
\ifx \showISSN     \undefined \def \showISSN      #1{\unskip}     \fi
\ifx \showLCCN     \undefined \def \showLCCN      #1{\unskip}     \fi
\ifx \shownote     \undefined \def \shownote      #1{#1}          \fi
\ifx \showarticletitle \undefined \def \showarticletitle #1{#1}   \fi
\ifx \showURL      \undefined \def \showURL       {\relax}        \fi
\providecommand\bibfield[2]{#2}
\providecommand\bibinfo[2]{#2}
\providecommand\natexlab[1]{#1}
\providecommand\showeprint[2][]{arXiv:#2}

\bibitem[\protect\citeauthoryear{Althoff, Danescu-Niculescu-Mizil, and
  Jurafsky}{Althoff et~al\mbox{.}}{2014}]%
        {althoff2014ask}
\bibfield{author}{\bibinfo{person}{Tim Althoff}, \bibinfo{person}{Cristian
  Danescu-Niculescu-Mizil}, {and} \bibinfo{person}{Dan Jurafsky}.}
  \bibinfo{year}{2014}\natexlab{}.
\newblock \showarticletitle{How to ask for a favor: A case study on the success
  of altruistic requests}. In \bibinfo{booktitle}{\emph{Eighth International
  AAAI Conference on Weblogs and Social Media}}.
\newblock


\bibitem[\protect\citeauthoryear{Barasinska and Sch{\"a}fer}{Barasinska and
  Sch{\"a}fer}{2014}]%
        {barasinska2014crowdfunding}
\bibfield{author}{\bibinfo{person}{Nataliya Barasinska} {and}
  \bibinfo{person}{Dorothea Sch{\"a}fer}.} \bibinfo{year}{2014}\natexlab{}.
\newblock \showarticletitle{Is crowdfunding different? Evidence on the relation
  between gender and funding success from a German peer-to-peer lending
  platform}.
\newblock \bibinfo{journal}{\emph{German Economic Review}}
  \bibinfo{volume}{15}, \bibinfo{number}{4} (\bibinfo{year}{2014}),
  \bibinfo{pages}{436--452}.
\newblock


\bibitem[\protect\citeauthoryear{Ceyhan, Shi, and Leskovec}{Ceyhan
  et~al\mbox{.}}{2011}]%
        {ceyhan2011dynamics}
\bibfield{author}{\bibinfo{person}{Simla Ceyhan}, \bibinfo{person}{Xiaolin
  Shi}, {and} \bibinfo{person}{Jure Leskovec}.}
  \bibinfo{year}{2011}\natexlab{}.
\newblock \showarticletitle{Dynamics of bidding in a P2P lending service:
  effects of herding and predicting loan success}. In
  \bibinfo{booktitle}{\emph{Proceedings of the 20th international conference on
  World wide web}}. ACM, \bibinfo{pages}{547--556}.
\newblock


\bibitem[\protect\citeauthoryear{Chatterjee and Barcun}{Chatterjee and
  Barcun}{1970}]%
        {chatterjee1970nonparametric}
\bibfield{author}{\bibinfo{person}{Samprit Chatterjee} {and}
  \bibinfo{person}{Seymour Barcun}.} \bibinfo{year}{1970}\natexlab{}.
\newblock \showarticletitle{A nonparametric approach to credit screening}.
\newblock \bibinfo{journal}{\emph{Journal of the American statistical
  Association}} \bibinfo{volume}{65}, \bibinfo{number}{329}
  (\bibinfo{year}{1970}), \bibinfo{pages}{150--154}.
\newblock


\bibitem[\protect\citeauthoryear{Drucker, Burges, Kaufman, Smola, and
  Vapnik}{Drucker et~al\mbox{.}}{1997}]%
        {drucker1997support}
\bibfield{author}{\bibinfo{person}{Harris Drucker},
  \bibinfo{person}{Christopher~JC Burges}, \bibinfo{person}{Linda Kaufman},
  \bibinfo{person}{Alex~J Smola}, {and} \bibinfo{person}{Vladimir Vapnik}.}
  \bibinfo{year}{1997}\natexlab{}.
\newblock \showarticletitle{Support vector regression machines}. In
  \bibinfo{booktitle}{\emph{Advances in neural information processing
  systems}}. \bibinfo{pages}{155--161}.
\newblock


\bibitem[\protect\citeauthoryear{Feldman and Gross}{Feldman and Gross}{2005}]%
        {feldman2005mortgage}
\bibfield{author}{\bibinfo{person}{David Feldman} {and}
  \bibinfo{person}{Shulamith Gross}.} \bibinfo{year}{2005}\natexlab{}.
\newblock \showarticletitle{Mortgage default: classification trees analysis}.
\newblock \bibinfo{journal}{\emph{The Journal of Real Estate Finance and
  Economics}} \bibinfo{volume}{30}, \bibinfo{number}{4} (\bibinfo{year}{2005}),
  \bibinfo{pages}{369--396}.
\newblock


\bibitem[\protect\citeauthoryear{Guo, Zhou, Luo, Liu, and Xiong}{Guo
  et~al\mbox{.}}{2016}]%
        {Guo2016417}
\bibfield{author}{\bibinfo{person}{Yanhong Guo}, \bibinfo{person}{Wenjun Zhou},
  \bibinfo{person}{Chunyu Luo}, \bibinfo{person}{Chuanren Liu}, {and}
  \bibinfo{person}{Hui Xiong}.} \bibinfo{year}{2016}\natexlab{}.
\newblock \showarticletitle{{Instance-based credit risk assessment for
  investment decisions in P2P lending}}.
\newblock \bibinfo{journal}{\emph{{European Journal of Operational Research}}}
  \bibinfo{volume}{249}, \bibinfo{number}{2} (\bibinfo{year}{2016}),
  \bibinfo{pages}{417 -- 426}.
\newblock
\urldef\tempurl%
\url{https://doi.org/10.1016/j.ejor.2015.05.050}
\showDOI{\tempurl}


\bibitem[\protect\citeauthoryear{Herzenstein, Andrews, Dholakia, and
  Lyandres}{Herzenstein et~al\mbox{.}}{2008}]%
        {herzenstein2008democratization}
\bibfield{author}{\bibinfo{person}{Michal Herzenstein}, \bibinfo{person}{Rick~L
  Andrews}, \bibinfo{person}{Utpal~M Dholakia}, {and} \bibinfo{person}{Evgeny
  Lyandres}.} \bibinfo{year}{2008}\natexlab{}.
\newblock \showarticletitle{The democratization of personal consumer loans?
  Determinants of success in online peer-to-peer lending communities}.
\newblock \bibinfo{journal}{\emph{Boston University School of Management
  Research Paper}} \bibinfo{volume}{14}, \bibinfo{number}{6}
  (\bibinfo{year}{2008}), \bibinfo{pages}{1--36}.
\newblock


\bibitem[\protect\citeauthoryear{Herzenstein, Dholakia, and
  Andrews}{Herzenstein et~al\mbox{.}}{2011a}]%
        {herzenstein2011strategic}
\bibfield{author}{\bibinfo{person}{Michal Herzenstein},
  \bibinfo{person}{Utpal~M Dholakia}, {and} \bibinfo{person}{Rick~L Andrews}.}
  \bibinfo{year}{2011}\natexlab{a}.
\newblock \showarticletitle{Strategic herding behavior in peer-to-peer loan
  auctions}.
\newblock \bibinfo{journal}{\emph{Journal of Interactive Marketing}}
  \bibinfo{volume}{25}, \bibinfo{number}{1} (\bibinfo{year}{2011}),
  \bibinfo{pages}{27--36}.
\newblock


\bibitem[\protect\citeauthoryear{Herzenstein, Sonenshein, and
  Dholakia}{Herzenstein et~al\mbox{.}}{2011b}]%
        {herzenstein2011tell}
\bibfield{author}{\bibinfo{person}{Michal Herzenstein}, \bibinfo{person}{Scott
  Sonenshein}, {and} \bibinfo{person}{Utpal~M Dholakia}.}
  \bibinfo{year}{2011}\natexlab{b}.
\newblock \showarticletitle{Tell me a good story and I may lend you money: The
  role of narratives in peer-to-peer lending decisions}.
\newblock \bibinfo{journal}{\emph{Journal of Marketing Research}}
  \bibinfo{volume}{48}, \bibinfo{number}{SPL} (\bibinfo{year}{2011}),
  \bibinfo{pages}{S138--S149}.
\newblock


\bibitem[\protect\citeauthoryear{Hutto and Gilbert}{Hutto and Gilbert}{2014}]%
        {hutto2014vader}
\bibfield{author}{\bibinfo{person}{Clayton~J Hutto} {and} \bibinfo{person}{Eric
  Gilbert}.} \bibinfo{year}{2014}\natexlab{}.
\newblock \showarticletitle{Vader: A parsimonious rule-based model for
  sentiment analysis of social media text}. In \bibinfo{booktitle}{\emph{Eighth
  international AAAI conference on weblogs and social media}}.
\newblock


\bibitem[\protect\citeauthoryear{Malekipirbazari and Aksakalli}{Malekipirbazari
  and Aksakalli}{2015}]%
        {malekipirbazari2015risk}
\bibfield{author}{\bibinfo{person}{Milad Malekipirbazari} {and}
  \bibinfo{person}{Vural Aksakalli}.} \bibinfo{year}{2015}\natexlab{}.
\newblock \showarticletitle{Risk assessment in social lending via random
  forests}.
\newblock \bibinfo{journal}{\emph{Expert Systems with Applications}}
  \bibinfo{volume}{42}, \bibinfo{number}{10} (\bibinfo{year}{2015}),
  \bibinfo{pages}{4621--4631}.
\newblock
\urldef\tempurl%
\url{https://doi.org/10.1016/j.eswa.2015.02.001}
\showDOI{\tempurl}


\bibitem[\protect\citeauthoryear{Neter, Kutner, Nachtsheim, and
  Wasserman}{Neter et~al\mbox{.}}{1996}]%
        {neter1996applied}
\bibfield{author}{\bibinfo{person}{John Neter}, \bibinfo{person}{Michael~H
  Kutner}, \bibinfo{person}{Christopher~J Nachtsheim}, {and}
  \bibinfo{person}{William Wasserman}.} \bibinfo{year}{1996}\natexlab{}.
\newblock \bibinfo{booktitle}{\emph{Applied linear statistical models}}.
  Vol.~\bibinfo{volume}{4}.
\newblock \bibinfo{publisher}{Irwin Chicago}.
\newblock


\bibitem[\protect\citeauthoryear{Prosper}{Prosper}{2018}]%
        {Prop}
\bibfield{author}{\bibinfo{person}{Prosper}.} \bibinfo{year}{2018}\natexlab{}.
\newblock \bibinfo{title}{Prosper Marketplace}.
\newblock \bibinfo{howpublished}{https://www.prosper.com/invest}.
\newblock
\newblock
\shownote{last accessed - 7/4/2019.}


\bibitem[\protect\citeauthoryear{Ren and Malik}{Ren and Malik}{2019}]%
        {ren2019investment}
\bibfield{author}{\bibinfo{person}{Ke Ren} {and} \bibinfo{person}{Avinash
  Malik}.} \bibinfo{year}{2019}\natexlab{}.
\newblock \showarticletitle{Investment Recommendation System for Low-Liquidity
  Online Peer to Peer Lending (P2PL) Marketplaces}. In
  \bibinfo{booktitle}{\emph{Proceedings of the Twelfth ACM International
  Conference on Web Search and Data Mining}}. ACM, \bibinfo{pages}{510--518}.
\newblock


\bibitem[\protect\citeauthoryear{Ryan, Reuk, and Wang}{Ryan
  et~al\mbox{.}}{2007}]%
        {ryan2007fund}
\bibfield{author}{\bibinfo{person}{Joe Ryan}, \bibinfo{person}{Katya Reuk},
  {and} \bibinfo{person}{Charles Wang}.} \bibinfo{year}{2007}\natexlab{}.
\newblock \showarticletitle{To fund or not to fund: Determinants of loan
  fundability in the prosper. com marketplace}.
\newblock \bibinfo{journal}{\emph{WP, The Standord Graduate School of
  Business}} (\bibinfo{year}{2007}).
\newblock


\bibitem[\protect\citeauthoryear{Xu}{Xu}{2015}]%
        {Udacity}
\bibfield{author}{\bibinfo{person}{Joash Xu}.} \bibinfo{year}{2015}\natexlab{}.
\newblock \bibinfo{title}{Prosper Loan Data}.
\newblock \bibinfo{howpublished}{https://github.com/joashxu/prosper-loan-data}.
\newblock
\newblock
\shownote{last accessed - 15/9/2016.}


\end{thebibliography}
